\documentclass[prd,amsmath,amssymb,superscriptaddress,preprintnumbers,twocolumn,nofootinbib,10pt]{revtex4-1}
\pdfoutput=1
\usepackage{graphicx}
\usepackage{dcolumn}
\usepackage{bm}
\usepackage{amssymb}
\usepackage{latexsym}
\usepackage{booktabs}
\usepackage{amsmath}
\usepackage{multirow}
\usepackage{url}
\usepackage{footnote}
\usepackage{float}
\usepackage{threeparttable}
\usepackage[colorlinks=true, linkcolor=blue, citecolor=blue]{hyperref}
\usepackage[bottom]{footmisc}

\usepackage[normalem]{ulem}
\usepackage{color}
\usepackage{array}
\usepackage{enumerate}
\usepackage{adjustbox}
\usepackage{makecell}
\usepackage{diagbox}
\usepackage{epstopdf}
\usepackage{epsfig}
\usepackage{longtable}
\usepackage{supertabular}
\usepackage{algorithm}
\usepackage{pifont}
\usepackage{algorithmic}
\usepackage{changepage}
\usepackage{setspace}

\begin{document}
\title{Impacts of dark energy on weighing neutrinos after DESI BAO}
\author{Guo-Hong Du}
\affiliation{Liaoning Key Laboratory of Cosmology and Astrophysics, College of Sciences, Northeastern University, Shenyang 110819, China}
\author{Peng-Ju Wu}
\affiliation{School of Physics, Ningxia University, Yinchuan 750021, China}
\author{Tian-Nuo Li}
\affiliation{Liaoning Key Laboratory of Cosmology and Astrophysics, College of Sciences, Northeastern University, Shenyang 110819, China}
\author{Xin Zhang}\thanks{Corresponding author}\email{zhangxin@mail.neu.edu.cn}
\affiliation{Liaoning Key Laboratory of Cosmology and Astrophysics, College of Sciences, Northeastern University, Shenyang 110819, China}
\affiliation{MOE Key Laboratory of Data Analytics and Optimization for Smart Industry, Northeastern University, Shenyang 110819, China}
\affiliation{National Frontiers Science Center for Industrial Intelligence and Systems Optimization, Northeastern University, Shenyang 110819, China}

\begin{abstract}
Recently, DESI has released baryon acoustic oscillation (BAO) data, and DES has also published its 5-year supernova (SN) data. These observations, combined with cosmic microwave background (CMB) data, support a dynamically evolving dark energy at a high confidence level. When using cosmological observations to weigh neutrinos, the results will be significantly affected by the measurement of dark energy due to the degeneracy between neutrino mass and the dark-energy equation of state. Therefore, we need to understand how the dynamical evolution of dark energy in the current situation will affect the measurement of neutrino mass. In this work, we utilize these latest observations and other additional distance measurements to discuss the mutual influence between neutrinos and dark energy, then calculate the Bayes factor to compare models. We consider three neutrino mass hierarchies, namely degenerate hierarchy (DH), normal hierarchy (NH), and inverted hierarchy (IH), as well as three dark energy models including $\Lambda \rm CDM$, $w\rm CDM$, and $w_0w_a \rm CDM$ models. We find that cosmological data combined with the prior of particle physics experiments can provide strong to decisive evidence favoring the $w_0w_a {\rm CDM}+\sum m_\nu$ model with NH. In the $w_0w_a \rm CDM$ model, using the CMB+DESI+DESY5 data, we obtain constraints on the total neutrino mass, $\sum m_\nu<0.171~\rm eV,~0.204~\rm eV,~0.220~\rm eV$, for DH, NH, and IH, respectively. Furthermore, taking into account the neutrino hierarchy or incorporating additional distance measurements results in a more pronounced deviation from the $\Lambda$CDM model for dark energy. The latter, in particular, exhibits a deviation at a confidence level that surpasses $4\sigma.$
\end{abstract}

\maketitle
\section{Introduction}
	
The phenomenon of neutrino oscillation indicates that neutrinos have mass, which is also the only confirmed new physics beyond the standard model of particle physics. The mass eigenstates of neutrinos are different from their flavor eigenstates, which is the root of the neutrino oscillation phenomenon. We denote the three mass eigenstates of neutrinos with $m_1$, $m_2$, and $m_3.$ Neutrino oscillation experiments have provided two mass-squared differences \cite{PhysRevLett.81.1562-Oscillation, PhysRevLett.89.011301-Oscillation, Olive_2014-Oscillation, dayabaycollaboration2022precision}, $\Delta m_{12}^2 = 7.54\times10^{-5}~\rm eV^2$ derived from solar and reactor neutrino observations and $\left|\Delta m_{32}^2\right| = 2.45\times10^{-3}~\rm eV^2$ from atmospheric and reactor neutrino experiments. 
These results indicate the existence of at least two massive neutrinos. However, the unknown sign of $\Delta m_{32}^2$ necessitates consideration of two possible mass hierarchies: normal ($\Delta m_{32}^2>0$) and inverted ($\Delta m_{32}^2<0$). 
In the normal hierarchy (NH), the two neutrinos that have the smallest mass splitting are lighter, requiring the total neutrino mass $\sum m_\nu \gtrsim 0.059~\rm eV.$ 
Conversely, in the inverted hierarchy (IH), it is the two heavier neutrinos that have the smallest mass splitting, yielding $\sum m_\nu \gtrsim 0.10~\rm eV$ \cite{deSalas:2020pgw-NHIH, Esteban:2020cvm-NHIH}.
In particle physics, the primary methods for measuring neutrino masses and confirming hierarchies are $\beta$-decay, neutrinoless double $\beta$-decay, and neutrino oscillation experiments \cite{BORNSCHEIN200514-Katrin,PhysRevLett.117.082503-0nu2beta,An_2016-JUNO,JUNO:2024jaw}. 
The latest $\beta$-decay experiment, conducted by the KATRIN collaboration in 2021, has achieved an effective mass measurement at the sub-electronvolt scale with $m_{\beta}<0.8~\rm eV$ \cite{aker2021direct-Katrin}. 
However, current measurements have provided only a relatively high upper limit for the mass of the lightest neutrino, and the precision has not yet reached the level required to distinguish between two hierarchies.

Cosmology provides an alternative approach to measuring neutrino masses, independent of particle physics. 
The hot big bang model predicts cosmic neutrino background (${\rm C}\nu\rm B$). 
It has not been directly observed but leaves detectable signatures in various cosmological measurements \cite{Lesgourgues_2012-neucosmo,article-neucosmo,Zhang:2017ljh,PTOLEMY:2019hkd,Vagnozzi:2019utt-neu,DiValentino:2024xsv-neucos}. 
In the early universe, neutrinos behaved as radiation, contributing to the total energy density and thus influencing the expansion history of the universe during the radiation-dominated era.
This influence can be studied through observations of big bang nucleosynthesis (BBN) and the integrated Sachs-Wolfe (ISW) effect in the cosmic microwave background (CMB).
At late times of the universe, neutrinos became non-relativistic and contributed to the matter component. 
We can explore the effect of neutrinos on the expansion history of the universe with low redshift probes such as baryon acoustic oscillations (BAO) and type Ia supernovae (SN Ia).
Furthermore, the relatively high velocities of neutrinos suppress the growth of small-scale structures, which are reflected in the power spectra of CMB and large-scale structures \cite{Zhang:2014dxk,Zhang:2014nta,Abazajian_2015-neuCMB,Coulton_2019-neuCMB,Zhang:2019ipd-21cmneu,Tanseri:2022zfe}.

In cosmology, when treating the total neutrino mass $\sum m_\nu$ as a free parameter, the degenerate hierarchy (DH) is usually considered, where all three neutrino masses are assumed to be identical \cite{LESGOURGUES2006307-DH}. 
This assumption can approximate the observable effects of both NH and IH to some extent. By appropriately adjusting the prior on $\sum m_\nu$, we can recover the NH and IH scenarios \cite{Valentino_2018-NHIH,Choudhury_2020-NHIH}. 
Cosmological observations are sensitive to the total neutrino mass \cite{desicollaboration2016desi}. For individual CMB observations, the latest constraint is $\sum m_\nu<0.24~\rm eV$ at the 95.4\% confidence level under the simplest $\Lambda$ cold dark matter ($\Lambda \rm CDM$) model with DH \cite{refId0-p18cosmo}.  

The use of cosmological observations to weigh neutrinos has its greatest limitation in that the weighing results are closely related to the cosmological model. This is because the neutrino mass is degenerate with other cosmological parameters and is difficult to measure independently in cosmology. In particular, the nature of dark energy will significantly affect the measurement results of neutrino mass 
\cite{Li:2012spm-neu-de,Zhang:2014ifa-neu-de,Zhang:2015rha-neu-de,Zhang_2016-neu-de,Feng:2017mfs-neu-de,Guo:2017hea-neu-de,Zhao:2016ecj-neu-de,Wang:2016tsz-neu-de,Zhao:2017urm-neu-de,Yang_2017-neu-de,Feng:2017nss-neu-de,Zhang:2017rbg-neu-de,Feng:2017usu-neu-de,Guo:2018gyo-neu-de,Choudhury_2018-neu-de,Zhang_2020-neu-de,Choudhury_2020-neu-de,Li:2020gtk-neu-de,Feng:2019jqa-neu-de,Feng:2019mym-neu-de,Zhao:2018fjj-neu-de,Feng:2021ipq-neu-de,Reeves:2022aoi-neu-de,Yao_2023-de}. 
Previous studies have reached some interesting conclusions. For example, in dynamical dark energy models, compared to $\Lambda$CDM, the upper limit of $\sum m_\nu$ can become larger and can also become smaller. It is found that, in the cases of phantom and early phantom (i.e., the quintom evolving from $w < -1$ to $w > -1$), the constraint on $\sum m_\nu$ becomes looser; but in the cases of quintessence and early quintessence (i.e., the quintom evolving from $w > -1$ to $w < -1$), the constraint on $\sum m_\nu$ becomes tighter \cite{Zhang:2017rbg-neu-de}. Here, $w\equiv p_{\rm de}/\rho_{\rm de}$ is the equation of state (EoS) of dark energy, with $p_{\rm de}$ and $\rho_{\rm de}$ the pressure and energy density of dark energy, respectively.

Recently, the Dark Energy Spectroscopic Instrument (DESI) presented the latest cosmological results obtained from the analysis of galaxy, quasar, and Lyman-$\alpha$ (Ly$\alpha$) forest BAO observations using data from the first year of observations \cite{DESI:2024uvr,DESI:2024lzq,DESI:2024mwx}. The DESI BAO data combined with CMB observations give the constraints on the neutrino mass, $\sum m_\nu<0.072$ eV, 0.113 eV, and 0.145 eV, for DH, NH, and IH, respectively, in $\Lambda \rm CDM$ \cite{DESI:2024mwx}.

In addition to the data release of DESI, as a component of the Year 5 data release, the Dark Energy Survey (DES) collaboration has released their SN Ia dataset, denoted as DESY5. Recent results from CMB, DESI, and DESY5 indicate that dark energy may be dynamical. 
The deviation from a cosmological constant $\Lambda$ for dark energy is at a confidence level of up to $3.9\sigma$ \cite{DESI:2024mwx}. 
This deviation has prompted extensive discourse on dark energy and its associated phenomena (see, e.g., Refs.~\cite{Tada:2024znt-desi, Giare:2024akf-desi, Pang:2024qyh, Reboucas:2024smm, Escamilla-Rivera:2024sae-desi,Pedrotti:2024kpn,park2024usingnondesidataconfirm,Wang:2024hen-desi, Li:2024qso-desi, Jiang:2024viw-desi,Ge:2024kac, Wu:2024faw,Chan-GyungPark:2024brx, Li:2024qus,Li:2025owk,Du:2025iow}).
Therefore, considering the strong degeneracy between neutrino mass and dark-energy EoS, it is necessary to explore the mutual influence between dark-energy EoS and neutrino mass using the current cosmological data.

In this work, we use current cosmological data from CMB, BAO, and SN Ia observations to derive upper limits on neutrino mass under different dark energy models and neutrino mass hierarchies. We also consider additional distance modulus data from gamma-ray bursts (GRBs) and angular diameter distance measurements of galaxy clusters for further analysis of neutrino mass. 
In addition, we calculate Bayesian evidence and compare Bayes factors for different models. 
We demonstrate the potential of cosmology to simultaneously determine the dark-energy EoS and neutrino mass. 
Finally, we perform a comparative analysis of the current cosmological results and particle physics experiment results.

The structure of this paper is as follows. In Sect. \ref{sec2}, we describe the theoretical analysis methods and datasets. In Sect. \ref{sec3}, we report and discuss the main results. Finally, we present conclusions of this paper in Sect. \ref{sec4}.

\section{Methodology}\label{sec2}

\subsection{Models}\label{sec2.1}

In cosmology, the total energy density of massive neutrinos (using the comoving momentum, $q$) is given by \cite{WMAP:2010qai}
\begin{equation}
\rho_\nu(a) = \frac{a^{-4}}{\pi^2} \int \frac{q^2\mathrm{d}q}{\mathrm{e}^{q/T_{\nu0}}+1}\sum_i \sqrt{q^2+m_{i}^2a^2},
\label{eq1}
\end{equation}
where $T_{\nu0}=(4/11)^{1/3}T_\mathrm{cmb}=1.945\,\mathrm{K}$ is the present neutrino temperature and $T_\mathrm{cmb}=2.725\,\mathrm{K}$ is the present CMB temperature. $m_i\,(i=1,2,3)$ represents the mass of each neutrino species.

In the early universe, neutrinos were relativistic, and therefore $\rho_{\nu}$ is proportional to the photon energy density, $\rho_{\gamma}$, given by \cite{WMAP:2010qai}
\begin{equation}
\rho_{\nu}(a) \to N_{\rm eff} \frac{7}{8} \left(\frac{4}{11}\right)^{4/3} \rho_\gamma(a),
\label{eq2}
\end{equation}
where $N_\mathrm{eff}=3.044$ is the effective number of relativistic neutrino species in the standard model \cite{Akita:2020szl, Froustey:2020mcq, Bennett:2020zkv}. Equation~(\ref{eq1}) can therefore be rewritten as
\begin{equation}
\rho_\nu(a) = N_{\rm eff} \frac{7}{8} \left(\frac{4}{11}\right)^{4/3} \rho_\gamma(a) \sum_i f\left(\frac{m_i}{T_{\nu0}}a\right),
\label{eq3}
\end{equation}
where 
\begin{equation}
f(y) \equiv \frac{40}{7\pi^4}\int_0^\infty \frac{x^2\sqrt{x^2+y^2}}{\mathrm{e}^x+1} \mathrm{d}x.
\label{eq4}
\end{equation}

In Eq.~(\ref{eq3}), we explicitly sum over three neutrino species due to the consideration of different neutrino hierarchies, and therefore Eq.~(\ref{eq4}) differs by a factor of 3 from Eq.~(25) in Ref.~\cite{WMAP:2010qai} which considers only the DH scenario. 
Due to the asymptotic behavior $3f(y)\to 1$ for $y\to 0$, Eq.~(\ref{eq3}) recovers Eq.~(\ref{eq2}) in the early-time limit $a\to 0.$ It can be demonstrated that when $a\to\infty$ (corresponding to the late-time universe), Eq.~(\ref{eq3}) tends asymptotically to
\begin{equation}
\rho_{\nu}(a) \to \frac{\sum m_{\nu}}{93.14h^2~\rm eV} \rho_\mathrm{crit,0}a^{-3},
\end{equation}
where $\rho_\mathrm{crit,0}=3H_0^2/8\pi G$ is the current critical density and $h=H_0/(100~\mathrm{km~s^{-1}~Mpc^{-1}})$ is the dimensionless Hubble constant. Utilizing the results derived from neutrino oscillation experiments, the total mass, $\sum m_\nu$, can be expressed as
\begin{equation}
	\sum m_\nu = m_1 + \sqrt{m_1^2 + \Delta m_{21}^2} + \sqrt{m_1^2 + \Delta m_{21}^2 + \left|\Delta m_{32}^2\right| }, 
\end{equation}	
for NH,
\begin{equation}	
	\sum m_\nu = m_3 + \sqrt{m_3^2 + \left|\Delta m_{32}^2\right| - \Delta m_{21}^2} + \sqrt{m_3^2 + \left|\Delta m_{32}^2\right|},
\end{equation}
for IH,
and $\sum m_\nu = 3m$, for DH. Meanwhile, we set the priors $\sum m_\nu > 0~\rm eV$ for DH, $\sum m_\nu > 0.06~\rm eV$ for NH, and $\sum m_\nu > 0.10~\rm eV$ for IH in the calculation.

Using the above results, in a flat Friedmann–Robertson–Walker universe, the Hubble parameter can be written as
\begin{widetext}
\begin{equation}
H(a) = H_0\sqrt{\Omega_\mathrm{m}a^{-3}+\Omega_\gamma a^{-4}\biggl[1+N_{\rm eff} \frac{7}{8} \left(\frac{4}{11}\right)^{4/3} \sum_i f\left(\frac{m_i}{T_{\nu0}}a\right)\biggr]+\Omega_\mathrm{de}a^{-3(1+w_\mathrm{eff}(a))}},
\end{equation}
\end{widetext}
where $\Omega_{\gamma}$, $\Omega_\mathrm{m}$, and $\Omega_{\rm de}$ represent the current density parameters of photon, matter, and dark energy, respectively, and we have $\Omega_\gamma = 2.469\times10^{-5}/h^2$ for $T_\mathrm{cmb}=2.725\,\mathrm{K}.$ Here, $w_\mathrm{eff}(a)=\frac{1}{\ln a}\int_0^{\ln a}w(a')\mathrm{d}\ln a'$ is the effective EoS of dark energy.

For dark energy, we consider three models, including the standard model $\Lambda {\rm CDM}+\sum m_\nu$ with $w=-1$, $w {\rm CDM}+\sum m_\nu$ model with a constant $w$, and $w_0w_a {\rm CDM}+\sum m_\nu$ model, which adopts a dynamical dark energy parameterization form with \cite{CHEVALLIER2001-w0wa,PhysRevLett.90.091301-w0wa}
\begin{equation}
	w(a) = w_0 + w_a(1-a).
\end{equation}

\subsection{Cosmological datasets}\label{sec2.2}

To analyze the neutrino sector in cosmology, we use the following datasets of cosmological observations.
\begin{itemize}
\item CMB. We use the CMB temperature power spectrum high-$\ell$ $plikTTTEEE$ \cite{efstathiou2019-highl, rosenberg22-highl}, low-$\ell$ $TT$ and $EE$ \cite{Pagano:2019tci-lowl} of Planck 2018 observations. 
Furthermore, we use the lensing data of the Atacama Cosmology Telescope (ACT)\footnote{\url{https://github.com/ACTCollaboration/act_dr6_lenslike}.} \cite{Madhavacheril_2024-actlen,Qu_2024-actlen} and Planck CMB PR4 (NPIPE) lensing data\footnote{\url{https://github.com/carronj/planck_PR4_lensing}.} \cite{Carron:2022eyg-PR4len, Carron:2022eum-PR4len}.
\item DESI. For DESI BAO observation, we use 12 BAO measurements in $0.1<z<4.16$\footnote{The DESI BAO data are from Data Release 1 (details at \url{https://data.desi.lbl.gov/doc/releases/}).} \cite{DESI:2024uvr, DESI:2024lzq, DESI:2024mwx}. 
This dataset in different redshift bins includes bright galaxy sample, luminous red galaxy sample, emission line galaxy sample, quasar sample, and Ly$\alpha$ forest sample.
\item PantheonPlus. We employ 1550 light curves with redshift range $0.01<z<2.26$ from spectroscopically confirmed SN compiled in the PantheonPlus dataset \cite{Brout:2022vxf-SN+}.
\item DESY5. We use 1829 photometrically categorized SN Ia data from the Year 5 data release of DES, with redshift range $0.025<z<1.3$ \cite{descollaboration2024dark-DESY5}.
\item LGRB. Here, LGRB stands for long gamma-ray bursts. LGRBs are associated with numerous astronomical phenomena, such as supernova explosions, and exhibit high redshifts due to high-energy characteristics, offering new possibilities for investigating neutrino mass. 
In our analysis, we utilize Hubble diagram data from 162 LGRBs in $0.03<z<9.3$ that have been calibrated using the supernova distances \cite{Demianski_2017-GRB}.
\item GADD. Here, GADD stands for galaxy cluster angular diameter distance. We incorporate angular diameter distance data for 25 galaxy clusters in $0.023<z<0.784$ under the triaxial ellipsoidal $\beta$-model assumption from X-ray and Sunyaev–Zeldovich observations \cite{De_Filippis_2005-GADD}.
\end{itemize}

\subsection{Bayesian analysis}\label{sec2.3}

We perform Bayesian inference on the set of basic cosmological parameters \{$\ln(10^{10} A_\mathrm{s})$, $n_\mathrm{s}$, $100\theta_\mathrm{MC}$, $\Omega_\mathrm{b} h^2$, $\Omega_\mathrm{c} h^2$, $\tau_\mathrm{reio}$, $\sum m_\nu$\} and dark-energy EoS parameters $w$ or \{$w_0$, $w_a$\}. 
We employ the Boltzmann code $\mathtt{CAMB}$\footnote{\url{https://github.com/cmbant/CAMB}.} \cite{Lewis:1999bs-camb,Howlett:2012mh-camb} for theoretical calculations in cosmology. Bayesian inference relies on the cosmological inference code $\mathtt{Cobaya}$\footnote{\url{https://github.com/CobayaSampler/cobaya}.} \cite{Torrado_2021-cobaya}, using Markov Chain Monte Carlo (MCMC) sampler \cite{Lewis:2002ah-mcmc,taking-mcmc,Lewis:2013hha-mcmc} with Metropolis-Hastings algorithm.
We stop the MCMC sampling when the Gelman–Rubin criterion \cite{10.1214/ss/1177011136-R-1} $R-1<0.01$ is fulfilled, assuming that the chains had converged. Moreover, $\mathtt{getdist}$\footnote{\url{https://github.com/cmbant/getdist}.} is used to analyze the results of the MCMC samples \cite{lewis2019getdist}.

Additionally, to quantitatively compare different neutrino hierarchy and dark energy models, we calculate the Bayes factor $\ln \mathcal{B}_{ij}$ in logarithmic space, given by
\begin{equation}
\ln \mathcal{B}_{ij} = \ln Z_i - \ln Z_j,
\label{eq: lnB}
\end{equation}
where $Z_i$ and $Z_j$ is Bayesian evidence of two models, which can be expressed as
\begin{equation}
Z \equiv P(D|M) = \int_{\Omega} P(D|\bm{\theta},M)P(\bm{\theta}|M)P(M)~{\rm d}\bm{\theta},
\label{eq: lnZ}
\end{equation}
where $P(D|M)$ is the probability of the data $D$ given the model $M$, $P(D|\bm{\theta},M)$ is the likelihood of $D$ given the parameters $\bm{\theta}$ and $M$, $P(\bm{\theta}|M)$ is the prior probability of $\bm{\theta}$ given $M$, and $P(M)$ is the prior of $M.$ 

It is worth noting that, in most cases, Bayesian analyses use the same prior $P(M)$ for the models being compared. However, in our analysis, we include the model prior from particle physics when comparing the NH and IH models, with $\ln P({\rm NH})/P({\rm IH}) = 6.5$ \cite{10.3389/fspas.2018.00036-lnNH/IH}. This choice reflects constraints from particle physics and is independent of the assumptions of dark energy models.

Typically, the Jeffreys scale \cite{Bayes-Jeffrey,Trotta_2008-Jeffrey} is employed to gauge the strength of model preference: if $\left|\ln \mathcal{B}_{ij}\right|<1$, the evidence is inconclusive; $1\le\left|\ln \mathcal{B}_{ij}\right|<2.5$ represents weak evidence; $2.5\le\left|\ln \mathcal{B}_{ij}\right|<5$ is moderate; $5\le\left|\ln \mathcal{B}_{ij}\right|<10$ is strong; and if $\left|\ln \mathcal{B}_{ij}\right|\ge 10$, the evidence is decisive.

Calculating Bayesian evidence is inherently challenging, particularly when it involves integrating over a high-dimensional parameter space, as is the case with the 16 to 18 dimensions in our study. To address this, we have adopted an approximation technique that relies on posterior samples. Specifically, we have utilized a machine learning-based normalizing flow approach that is grounded in the harmonic mean estimation. This method is implemented through the use of the $\mathtt{harmonic}$\footnote{\url{https://github.com/astro-informatics/harmonic}.} package, which has been referenced in several studies \cite{machine-harmonic,rzad051-harmonic,polanska2024learned-harmonic}.  

Like many machine learning techniques, the harmonic method has demonstrated robustness and the ability to generalize well. It has been shown to accurately model the posterior distributions within the 18-dimensional parameter space of our investigation, effectively matching the parameter samples provided. Moreover, although \texttt{Cobaya} provides nested sampling methods such as \texttt{PolyChord} for direct evidence computation, the harmonic method offers a significant computational advantage, achieving a speedup of approximately six times while maintaining consistency with nested sampling results \cite{polanska2024learned-harmonic}.

Given these strengths, we have chosen to apply the harmonic method to evaluate potential selection biases across different cosmological models in our analysis.

\section{Results and discussion}\label{sec3}

\begin{table*}[t]
\renewcommand\arraystretch{1.7}
\centering
\caption{The constraints on cosmological parameters from the current cosmological datasets. For comparison, the previous results from CMB+DR16+Pantheon \cite{DR16+Pantheon-Di_Valentino_2022} datasets are also listed. Note that in this paper, the CMB data include ACT and Planck PR4 lensing. The upper limit on \(\sum m_\nu\) represents the 95.4\% ($2\sigma$) confidence limit, while the results for other parameters are given as the 68.3\% ($1\sigma$) confidence intervals of the marginalized mean values.}
\label{table: cosmic results}
\resizebox{\textwidth}{!}{%
\large
\begin{tabular}{cccccccccc}
\toprule[1pt]
& \multicolumn{3}{c}{$~~\Lambda\mathrm{CDM}+\sum m_{\nu}~~$} & \multicolumn{3}{c}{$~w\mathrm{CDM}+\sum m_{\nu}~$} & \multicolumn{3}{c}{$w_0w_a\mathrm{CDM}+\sum m_{\nu}$} \\
\cmidrule[0.5pt](l{2pt}r{2pt}){2-4} \cmidrule[0.5pt](l{2pt}r{2pt}){5-7} \cmidrule[0.5pt](l{2pt}r{2pt}){8-10}
& DH & NH & IH & DH & NH & IH & DH & NH & IH \\
\midrule[1pt]
\multicolumn{10}{l}{\textbf{CMB+DR16+Pantheon}} \\
$\sum m_{\nu}(2\sigma)$ [eV] & $< 0.087$ & $< 0.129$ & $< 0.155$ & $< 0.139$ & $< 0.165$ & $< 0.204$ & $< 0.224$ & $< 0.248$ & $< 0.265$ \\
\midrule[0.8pt]
\multicolumn{10}{l}{\textbf{CMB+DESI+PantheonPlus}} \\
$H_0$ [km s$^{-1}$ Mpc$^{-1}$] & $67.92^{+0.40}_{-0.40}$ & $67.59^{+0.38}_{-0.38}$ & $67.34^{+0.37}_{-0.37}$ & $67.65^{+0.70}_{-0.70}$ & $67.58^{+0.69}_{-0.69}$ & $67.51^{+0.68}_{-0.68}$ & $67.94^{+0.71}_{-0.71}$ & $67.89^{+0.71}_{-0.71}$ & $67.84^{+0.70}_{-0.70}$ \\
$w/w_0$ & -- & -- & -- & $-0.988^{+0.026}_{-0.026}$ & $-0.999^{+0.026}_{-0.026}$ & $-1.007^{+0.025}_{-0.025}$ & $-0.828^{+0.066}_{-0.066}$ & $-0.816^{+0.065}_{-0.065}$ & $-0.810^{+0.066}_{-0.066}$ \\
$w_a$ & -- & -- & -- & -- & -- & -- & $-0.75^{+0.34}_{-0.26}$ & $-0.86^{+0.33}_{-0.27}$ & $-0.92^{+0.33}_{-0.28}$ \\
$\sum m_{\nu}(2\sigma)$ [eV] & $< 0.077$ & $< 0.125$ & $< 0.153$ & $< 0.076$ & $< 0.128$ & $< 0.158$ & $< 0.147$ & $< 0.181$ & $< 0.203$ \\
\midrule[0.8pt]
\multicolumn{10}{l}{\textbf{CMB+DESI+DESY5}} \\
$H_0$ [km s$^{-1}$ Mpc$^{-1}$] & $67.74^{+0.42}_{-0.39}$ & $67.41^{+0.39}_{-0.39}$ & $67.17^{+0.37}_{-0.37}$ & $66.80^{+0.64}_{-0.64}$ & $66.75^{+0.62}_{-0.62}$ & $66.71^{+0.63}_{-0.63}$ & $67.10^{+0.66}_{-0.66}$ & $67.01^{+0.67}_{-0.67}$ & $67.00^{+0.65}_{-0.65}$ \\
$w/w_0$ & -- & -- & -- & $-0.957^{+0.023}_{-0.023}$ & $-0.970^{+0.023}_{-0.023}$ & $-0.980^{+0.024}_{-0.024}$ & $-0.725^{+0.066}_{-0.074}$ & $-0.709^{+0.072}_{-0.072}$ & $-0.698^{+0.066}_{-0.076}$ \\
$w_a$ & -- & -- & -- & -- & -- & -- & $-1.08^{+0.38}_{-0.30}$ & $-1.19^{+0.37}_{-0.32}$ & $-1.28^{+0.38}_{-0.30}$ \\
$\sum m_{\nu}(2\sigma)$ [eV] & $< 0.091$ & $< 0.130$ & $< 0.156$ & $< 0.069$ & $< 0.118$ & $< 0.149$ & $< 0.171$ & $< 0.204$ & $< 0.220$ \\
\midrule[1pt]
\multicolumn{10}{l}{\textbf{CMB+DESI+PantheonPlus+LGRB+GADD}} \\
$H_0$ [km s$^{-1}$ Mpc$^{-1}$] & $68.55^{+0.34}_{-0.34}$ & $68.26^{+0.34}_{-0.34}$ & $68.05^{+0.33}_{-0.33}$ & $68.89^{+0.55}_{-0.55}$ & $68.88^{+0.56}_{-0.56}$ & $68.86^{+0.54}_{-0.54}$ & $68.74^{+0.55}_{-0.55}$ & $68.67^{+0.57}_{-0.57}$ & $68.64^{+0.56}_{-0.56}$ \\
$w/w_0$ & -- & -- & -- & $-1.018^{+0.022}_{-0.022}$ & $-1.031^{+0.022}_{-0.022}$ & $-1.041^{+0.022}_{-0.022}$ & $-0.804^{+0.064}_{-0.064}$ & $-0.792^{+0.064}_{-0.064}$ & $-0.782^{+0.064}_{-0.064}$ \\
$w_a$ & -- & -- & -- & -- & -- & -- & $-0.94^{+0.31}_{-0.26}$ & $-1.04^{+0.30}_{-0.27}$ & $-1.14^{+0.31}_{-0.27}$ \\
$\sum m_{\nu}(2\sigma)$ [eV] & $< 0.048$ & $< 0.099$ & $< 0.134$ & $< 0.056$ & $< 0.107$ & $< 0.136$ & $< 0.114$ & $< 0.158$ & $< 0.185$ \\
\midrule[0.8pt]
\multicolumn{10}{l}{\textbf{CMB+DESI+DESY5+LGRB+GADD}} \\
$H_0$ [km s$^{-1}$ Mpc$^{-1}$] & $68.39^{+0.33}_{-0.33}$ & $68.09^{+0.33}_{-0.33}$ & $67.88^{+0.32}_{-0.32}$ & $68.26^{+0.52}_{-0.52}$ & $68.22^{+0.53}_{-0.53}$ & $68.20^{+0.52}_{-0.52}$ & $67.97^{+0.53}_{-0.53}$ & $67.91^{+0.53}_{-0.53}$ & $67.87^{+0.53}_{-0.53}$ \\
$w/w_0$ & -- & -- & -- & $-0.993^{+0.020}_{-0.020}$ & $-1.007^{+0.021}_{-0.021}$ & $-1.017^{+0.021}_{-0.021}$ & $-0.698^{+0.069}_{-0.069}$ & $-0.680^{+0.070}_{-0.070}$ & $-0.667^{+0.068}_{-0.068}$ \\
$w_a$ & -- & -- & -- & -- & -- & -- & $-1.28^{+0.35}_{-0.29}$ & $-1.41^{+0.34}_{-0.29}$ & $-1.52^{+0.34}_{-0.30}$ \\
$\sum m_{\nu}(2\sigma)$ [eV] & $< 0.050$ & $< 0.099$ & $< 0.134$ & $< 0.048$ & $< 0.103$ & $< 0.139$ & $< 0.137$ & $< 0.166$ & $< 0.192$ \\
\bottomrule[1pt]
\end{tabular}
}
\end{table*}

\subsection{Results with current CMB+BAO+SN data}\label{sec3.1}

\begin{figure*}[t]
\resizebox{\textwidth}{!}{
\begin{minipage}{0.48\textwidth}
\centering
\includegraphics[width=\linewidth]{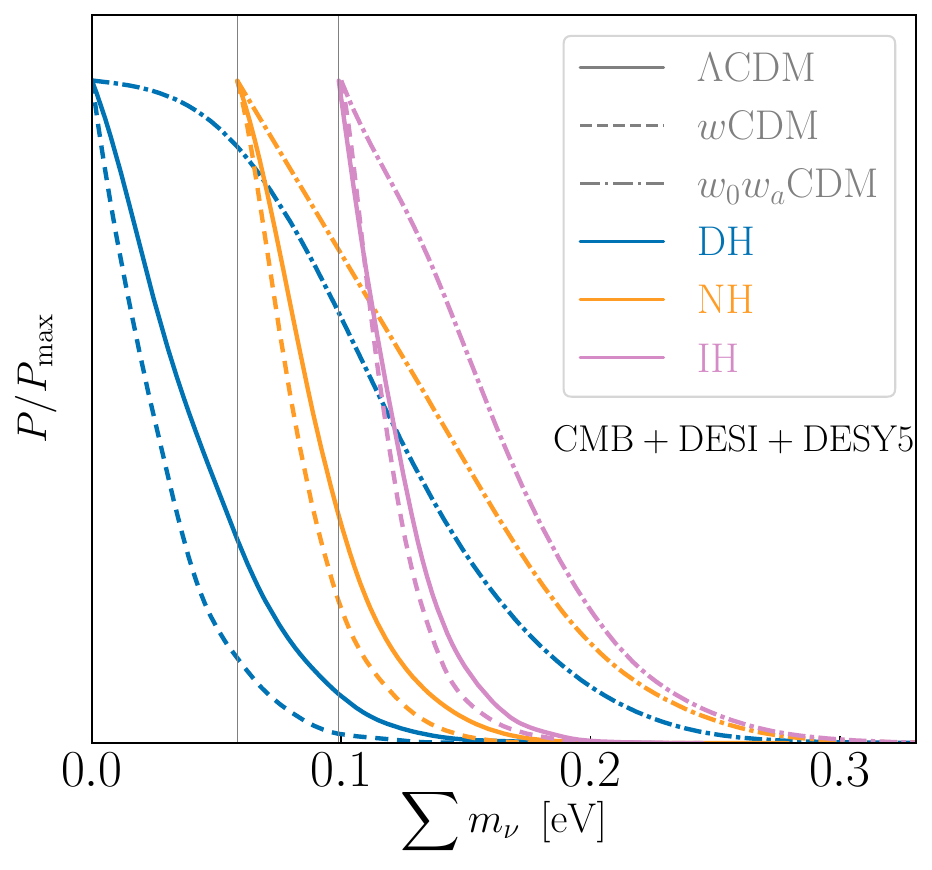}
\end{minipage}\hfill
\begin{minipage}{0.46\textwidth}
\centering
\includegraphics[width=\linewidth]{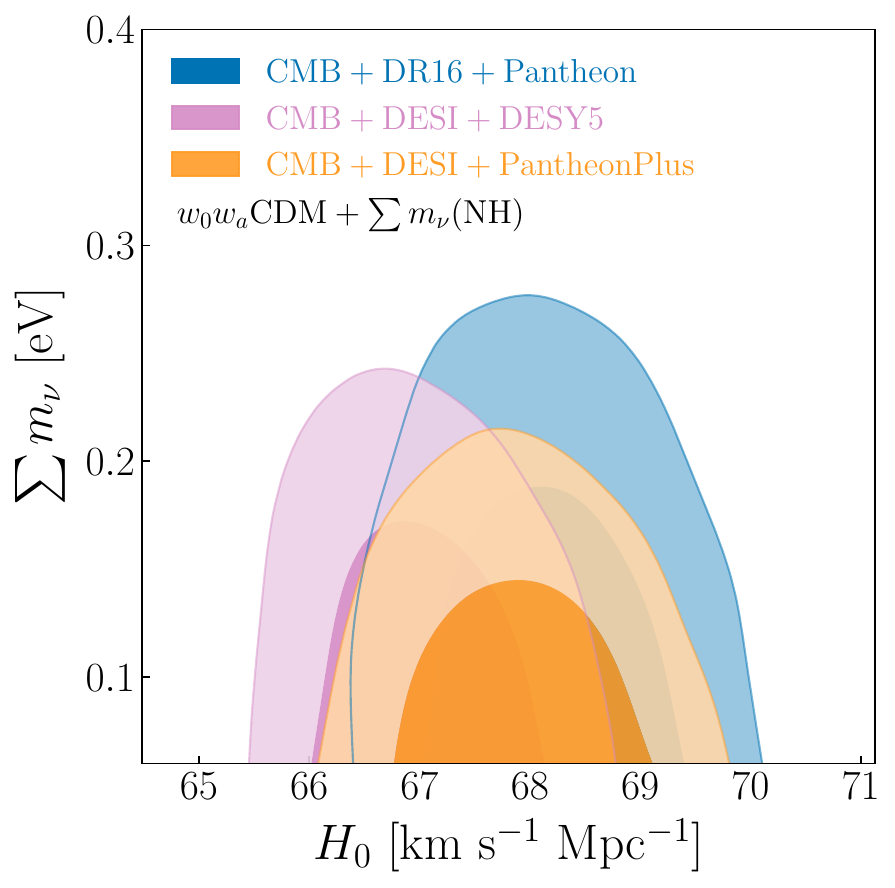}
\end{minipage}
\begin{minipage}{0.46\textwidth}
\centering
\includegraphics[width=\linewidth]{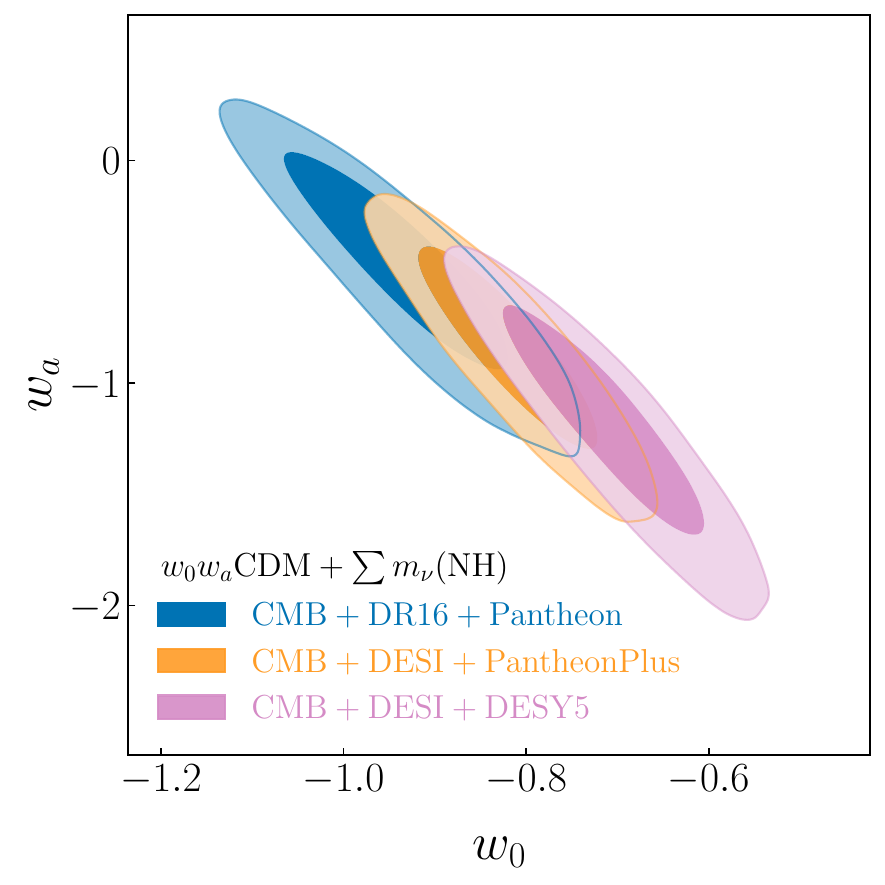}
\end{minipage}
}
\centering
\caption{The cosmological constraint results of CMB, BAO, and SN data. {\it Left panel}: The one-dimensional (1D) marginalized posterior constraints on $\sum m_\nu$ using the CMB+DESI+DESY5 datasets in different dark energy models and neutrino hierarchies. Solid line, dashed line, and dot-dashed line correspond to the $\Lambda\rm CDM$, $w\rm CDM$, and $w_0w_a\rm CDM$ models respectively, while blue, orange, and magenta represent DH, NH, and IH.
{\it Middle and right panels}: The 68.3\% and 95.4\% credible-interval contours for $\sum m_\nu$ and $H_0$, as well as $w_0$ and $w_a$ in the $w_0w_a{\rm CDM}+\sum m_\nu~(\rm NH)$ model, using the CMB+DR16+Pantheon, CMB+DESI+DESY5, and CMB+DESI+PantheonPlus datasets (marked in blue, magenta, and orange, respectively). DR16 refers to the BAO and redshift space distortion (RSD) measurements from the Sloan Digital Sky Survey (SDSS) spectroscopic galaxy and quasar catalog, including 6DF, DR7, BOSS DR12 and eBOSS DR16 \cite{Beutler:2012px-6DF,Ross:2014qpa-DR7,Alam:2016hwk-DR12,Alam:2020sor-DR16}. The Pantheon dataset compiles supernova data from 1048 SN samples \cite{Scolnic:2017caz-Pantheon}.}
\label{Fig: mnu}
\end{figure*}

We present the constraints on cosmological parameters in Table~\ref{table: cosmic results}, utilizing the current datasets CMB+DESI+PantheonPlus and CMB+DESI+DESY5. The upper limits for the sum of neutrino masses, $\sum m_\nu$, are reported at the 95.4\% confidence level, corresponding to 2$\sigma$, while constraints on other parameters are provided at the 68.3\% confidence level, corresponding to 1$\sigma.$

As detailed in Table~\ref{table: cosmic results}, the latest cosmological data have further lowered the upper limits of $\sum m_\nu$, particularly within the $w\rm CDM$ and $w_0w_a\rm CDM$ models. For instance, the CMB+DESI+PantheonPlus data yield $\sum m_\nu<0.147$ eV, 0.181 eV, and 0.203 eV for DH, NH, and IH in the $w_0w_a\rm CDM$ model, respectively. This represents a reduction in the upper bounds of $\sum m_\nu$ by 34.4\%, 27.0\%, and 23.4\%, respectively, compared to the previously established upper bounds using CMB+DR16+Pantheon data \cite{DR16+Pantheon-Di_Valentino_2022}. Similarly, within the $w\rm CDM$ model, the CMB+DESI+DESY5 datasets provide $\sum m_\nu<0.0659$ eV, 0.115 eV, and 0.152 eV for DH, NH, and IH, corresponding to reductions of 52.6\%, 30.3\%, and 25.5\%, respectively.

Figure~\ref{Fig: mnu} shows that the introduction of dynamically evolving dark energy leads to a significant loosening in the upper limit of $\sum m_\nu.$ The CMB+DESI+DESY5 data yield $\sum m_\nu<0.171$ eV, 0.204 eV, and 0.220 eV in the $w_0w_a{\rm CDM}+\sum m_\nu$ model for DH, NH, and IH, respectively. These results are highly consistent with the previous conclusion that when the EoS of dark energy evolves from $w<-1$ to $w>-1$, the constraint on $\sum m_\nu$ becomes looser \cite{Zhang:2017rbg-neu-de}.

Focusing on the CMB+DESI+DESY5 data, we observe that despite the $w\rm CDM$ model introducing an additional parameter $w$ compared to the $\Lambda \rm CDM$ model, the upper bound of $\sum m_\nu$ is actually tightened due to a lower value of $H_0$ and a higher value of $w$, which also accords closely with the previous study \cite{Feng:2019jqa-neu-de,Feng:2019mym-neu-de}. Additionally, the PantheonPlus dataset provides a more stringent upper limit for $\sum m_\nu$ than the DESY5 dataset. 

Furthermore, for the CMB+DESI+PantheonPlus data, we obtain ($w_0$, $w_a$) = ($-0.828^{+0.066}_{-0.066}$, $-0.75^{+0.34}_{-0.22}$), ($-0.816^{+0.065}_{-0.065}$, $-0.86^{+0.33}_{-0.27}$), and ($-0.810^{+0.066}_{-0.066}$, $-0.92^{+0.33}_{-0.28}$), while the CMB+DESI+DESY5 data yield ($w_0$, $w_a$) = ($-0.725^{+0.066}_{-0.074}$, $-1.08^{+0.38}_{-0.30}$), ($-0.709^{+0.072}_{-0.072}$, $ -1.19^{+0.37}_{-0.32}$), and ($-0.698^{+0.066}_{-0.076}$, $-1.28^{+0.38}_{-0.30}$) for DH, NH, and IH, respectively. 
These results indicate that when the effects of neutrino mass and hierarchy are taken into account, the deviation of dark energy behavior from the $\Lambda \rm CDM$ model becomes more pronounced. Consequently, although the precision of constraints may degrade due to the inclusion of neutrino mass, the behavior of dark energy still tends to align more closely with the $w_0w_a\rm CDM$ model. Using CMB+DESI+DESY5 data, this is evidenced by significance levels of $3.2\sigma$, $3.4\sigma$, and $3.8\sigma$ for DH, NH, and IH, respectively, consistent with that reported in Ref.~\cite{DESI:2024mwx}.

\subsection{Results with additional data}\label{sec3.2}

\begin{figure*}[t]
\resizebox{\textwidth}{!}{
\begin{minipage}{0.48\textwidth}
\centering
\includegraphics[width=\linewidth]{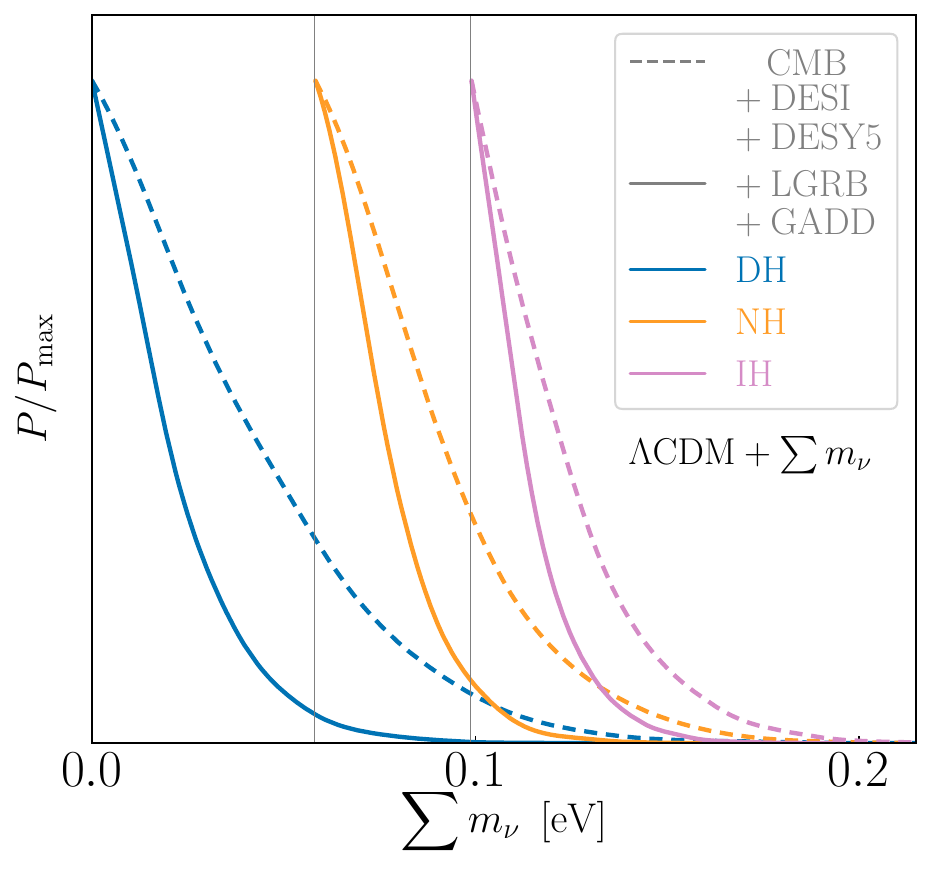}
\end{minipage}\hfill
\begin{minipage}{0.46\textwidth}
\centering
\includegraphics[width=\linewidth]{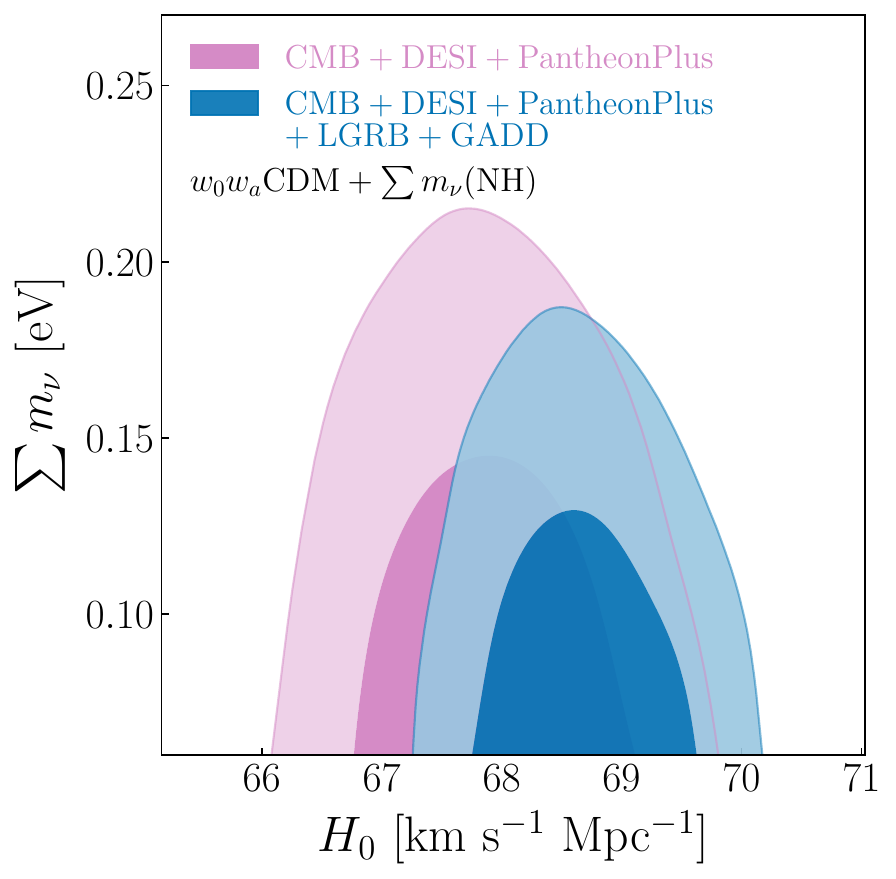}
\end{minipage}
\begin{minipage}{0.46\textwidth}
\centering
\includegraphics[width=\linewidth]{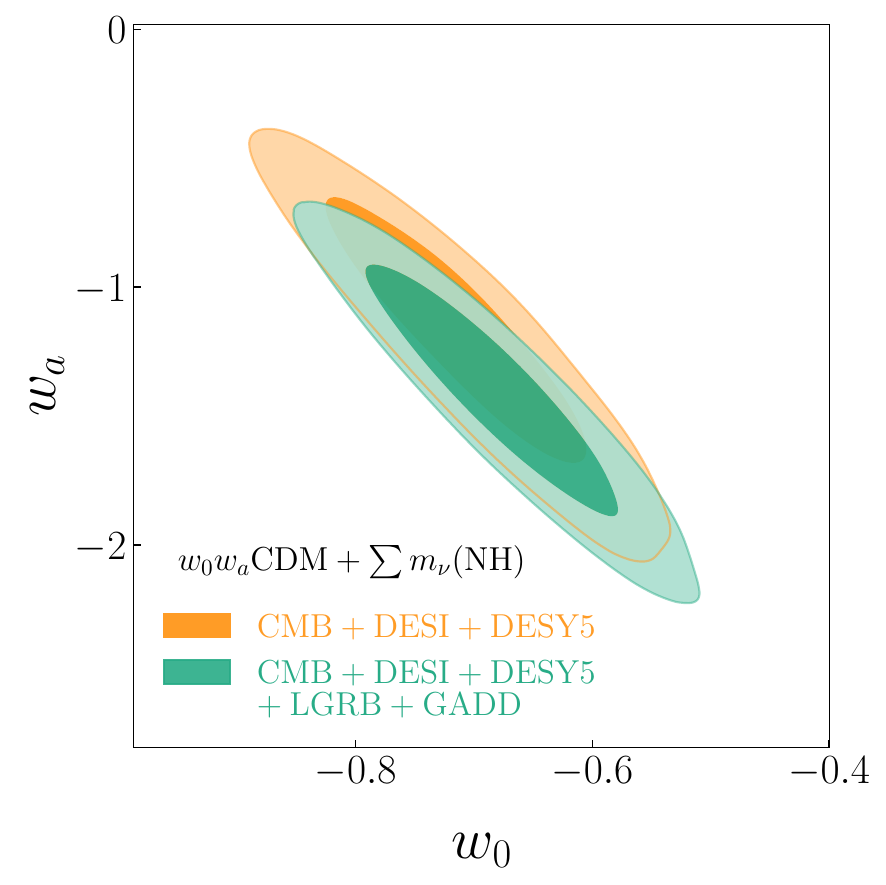}
\end{minipage}
}
\centering
\caption{The cosmological constraint results by adding two external distance observations. {\it Left panel}: The 1D marginalized posterior constraints on $\sum m_\nu$ using the CMB+DESI+DESY5 and CMB+DESI+DESY5+LGRB+GADD (shown as solid lines and dashed lines) datasets in $\Lambda{\rm CDM}+\sum m_\nu.$ Blue, orange, and magenta represent DH, NH, and IH. 
{\it Middle and right panel}: The 68.3\% and 95.4\% credible-interval contours for $\sum m_\nu$ and $H_0$, as well as $w_0$ and $w_a$ in the $w_0w_a{\rm CDM}+\sum m_\nu(\rm NH)$ model, using the CMB+DESI+PantheonPlus, and CMB+DESI+PantheonPlus+LGRB+GADD, CMB+DESI+DESY5, CMB+DESI+DESY5+LGRB+GADD datasets (marked in magenta, blue, orange, and green, respectively).}
\label{Fig: mnu2}
\end{figure*}

In this section, we extend our analysis by incorporating two additional sets of cosmological distance observations: LGRB and GADD. The results are detailed in Table~\ref{table: cosmic results} and depicted in Fig.~\ref{Fig: mnu2}.

Upon incorporating LGRB and GADD data, the upper limit of the sum of neutrino masses, $\sum m_\nu$, is somewhat reduced across various dark energy models and neutrino hierarchies. For instance, the PantheonPlus dataset, when compared to DESY5, provides more stringent constraints on $\sum m_\nu$, resulting in $\sum m_\nu<0.114$ eV, 0.158 eV, and 0.185 eV for DH, NH, and IH in the $w_0w_a$CDM model, respectively.

Furthermore, the datasets CMB+DESI+PantheonPlus+LGRB+GADD and CMB+DESI+DESY5+LGRB+GADD report $\sigma(H_0)=0.55~\rm km~s^{-1}~Mpc^{-1}$ and $\sigma(H_0)=0.53~\rm km~s^{-1}~Mpc^{-1}$, respectively. The precision of the Hubble constant $H_0$ constraints has improved by approximately 20\% when these additional datasets are included, as opposed to the scenario without LGRB and GADD. This indicates that the inclusion of external data from LGRB and GADD can effectively alleviate parameter degeneracy and yield a more precise constraint on $H_0.$

Additionally, the higher values of $H_0$ provided by GADD and LGRB contribute to the tighter neutrino mass limit. Moreover, the CMB+DESI+DESY5+LGRB+GADD data yield ($w_0$, $w_a$) = ($-0.698^{+0.069}_{-0.069}$, $-1.28^{+0.35}_{-0.29}$), ($-0.680^{+0.070}_{-0.070}$, $-1.41^{+0.34}_{-0.29}$), and ($-0.667^{+0.068}_{-0.068}$, $-1.52^{+0.34}_{-0.30}$) for DH, NH, and IH, respectively. This suggests that the inclusion of LGRB and GADD data leads to a more pronounced deviation of dark energy behavior from the $\Lambda \rm CDM$ model, with $4.0\sigma$, $4.4\sigma$, and $4.8\sigma$ significance levels for DH, NH, and IH, respectively, despite only a marginal improvement in the precision of the constraints on the dark-energy EoS parameters $w_0$ and $w_a.$

\subsection{Bayes factor}\label{sec3.3}

\begin{table*}[t]
\renewcommand\arraystretch{1.7}
\centering
\caption{The Bayes factors calculated with the harmonic method using cosmological datasets and particle prior. Note that $\ln \mathcal{B}_{ij}$ is computed as $Z_j$ in Eq.~(\ref{eq: lnB}) taken to be the Bayesian evidence of $w_0w_a{\rm CDM}+\sum m_\nu({\rm NH})$ model.}
\label{table: lnB}
\resizebox{0.6\textwidth}{!}{%
\begin{tabular}{cccccccccc}
\toprule[1pt]
& \multicolumn{3}{c}{$~~\Lambda\mathrm{CDM}+\sum m_{\nu}~~$} & \multicolumn{3}{c}{$~w\mathrm{CDM}+\sum m_{\nu}~$} & \multicolumn{3}{c}{$w_0w_a\mathrm{CDM}+\sum m_{\nu}$} \\
\cmidrule[0.5pt](l{2pt}r{2pt}){2-4} \cmidrule[0.5pt](l{2pt}r{2pt}){5-7} \cmidrule[0.5pt](l{2pt}r{2pt}){8-10}
& DH & NH & IH & DH & NH & IH & DH & NH & IH \\
\midrule[1pt]
\multicolumn{10}{l}{\small\textbf{CMB+DESI+PantheonPlus+LGRB+GADD}} \\
$\ln\mathcal{B}_{ij}$ & $0.102$ & $-3.050$ & $-6.057$ & $-3.524$ & $-6.367$ & $-8.578$ & $~0.920$ & $\quad0\quad$ & $-3.205$ \\
\midrule[0.8pt]
\multicolumn{10}{l}{\small\textbf{CMB+DESI+DESY5+LGRB+GADD}} \\
$\ln\mathcal{B}_{ij}$ & $-5.136$ & $-8.354$ & $-11.391$ & $-9.120$ & $-12.270$ & $-15.112$ & $0.773$ & $\quad0\quad$ & $-0.918$ \\
\midrule[0.8pt]
\multicolumn{10}{l}{\small\textbf{CMB+DESI+DESY5+LGRB+GADD+Particle}} \\
$\ln\mathcal{B}_{ij}$ & $-11.636$ & $-8.354$ & $-17.891$ & $-15.620$ & $-12.270$ & $-21.612$ & $-5.727$ & $\quad0\quad$ & $-7.418$ \\
\bottomrule[1pt]
\end{tabular}
}
\end{table*}

\begin{figure*}[t]
\resizebox{\textwidth}{!}{
\begin{minipage}{\textwidth}
\centering
\includegraphics[width=\linewidth]{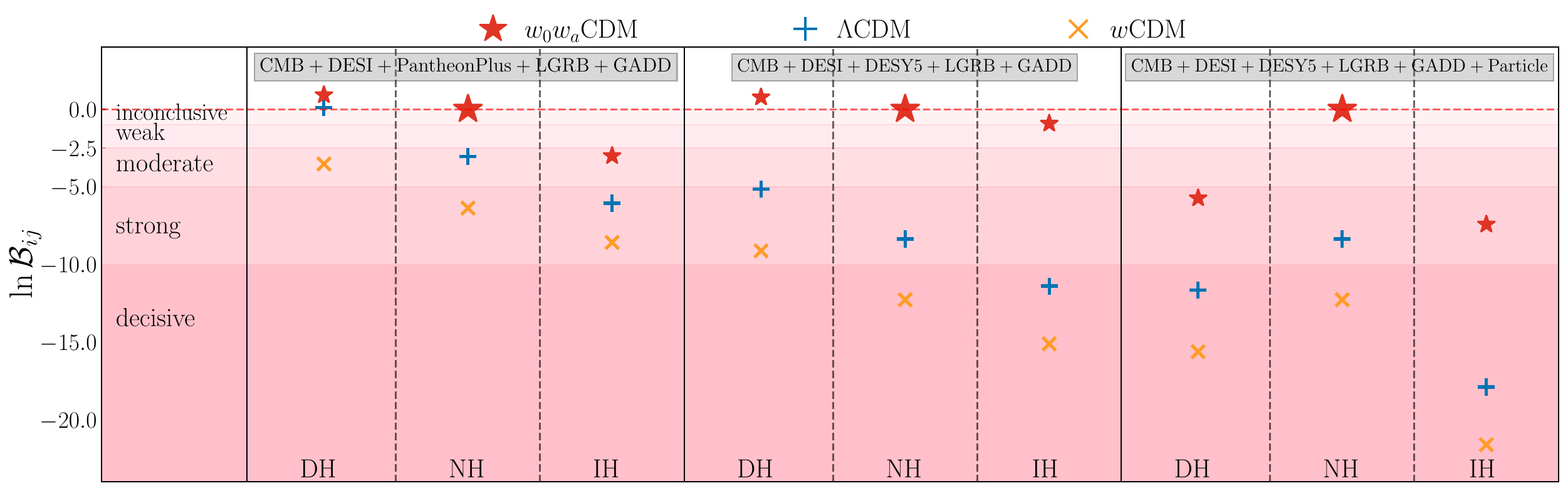}
\end{minipage}
}
\centering
\caption{The Bayes factors listed in Table~\ref{table: lnB}. We use red pentagrams, blue plus signs, and orange multiplication signs to represent \( w_0w_a \rm CDM\), $\Lambda \rm CDM$, and \( w\rm CDM \), respectively. According to the Jeffreys scale \cite{Trotta_2008-Jeffrey}, we use shades of pink from light to dark to indicate inconclusive, weak, moderate, strong, and decisive evidence, respectively. We mark the line $\ln \mathcal{B}_{ij} = 0$ with a red dashed line for the $w_0w_a$CDM+$\sum m_\nu$(NH) model.}
\label{Fig: Bayes}
\end{figure*}

The Bayes factors for various models, as derived using the harmonic method, are enumerated in Table~\ref{table: lnB}. Figure~\ref{Fig: Bayes} offers a visual representation of these results. It should be noted that all Bayes factors computed using different datasets are based on taking the Bayesian evidence of the $w_0w_a{\rm CDM}+\sum m_\nu~(\rm NH)$ model as the model $j$ in Eq.~(\ref{eq: lnB}). Therefore, a negative value of the Bayes factor $\ln \mathcal{B}_{ij}$ indicates that model $i$ is less supported by the observational data than model $j$; conversely, a positive value suggests that model $i$ is more supported by the data.


While Fig.~\ref{Fig: Bayes} indicates a tendency towards favoring DH across all three dark energy models, it is essential to recognize that the DH assumption, not aligned with particle physics findings, is only suitable for comparative analysis. Neutrino oscillation experiments have provided clear determinations of mass differences, necessitating that only NH and IH be considered in model comparisons. Within the context of the models examined in this study, the CMB+DESI+PantheonPlus+LGRB+GADD data offer moderate to strong evidence in support of the $w_0w_a \rm CDM$ model with NH. This data combination also robustly excludes the $w \rm CDM$ model and the $\Lambda \rm CDM$ model with IH, providing strong evidence to that effect. As shown in the middle panel of Fig.~\ref{Fig: Bayes}, the $\rm CMB+DESI+DESY5+LGRB+GADD$ data provide almost decisive evidence for the $w_0w_a$CDM model when comparing dark energy models under fixed neutrino mass hierarchy (either IH or NH). Conversely, when fixing the dark energy model to the $w_0w_a$CDM model, the comparison between neutrino mass hierarchies yields only inconclusive evidence for NH relative to IH. Thus, DESY5 demonstrates a greater capacity to distinguish between dark energy models than PantheonPlus, yet it is less effective in discerning neutrino mass hierarchies, as reflected in their respective cosmological outcomes.

Moreover, data from neutrino oscillation experiments impose a prior preference for the neutrino hierarchy, resulting in \( \ln \mathcal{B}_{\rm NH/IH} = 6.5 \) when considering only the central value \cite{10.3389/fspas.2018.00036-lnNH/IH}. This implies that for the model prior  \( P(M) \) in Eq.~(\ref{eq: lnZ}), the difference \( \ln P({\rm NH})-\ln P({\rm IH}) \) equals 6.5, irrespective of the dark energy model. Therefore, by integrating the latest cosmological data, such as CMB+DESI+DESY5+LGRB+GADD, with priors from particle physics experiments, we can achieve strong to decisive evidence favoring the $w_0w_a\mathrm{CDM}$ model with NH. As a result, cosmology and particle physics complement each other, offering the ability to effectively distinguish between dark energy models and neutrino hierarchies, while also imposing stricter constraints on the upper limit of the total neutrino mass.

\subsection{Comparison with particle physics experiments}\label{sec3.4}

\begin{figure*}[t]
\resizebox{\textwidth}{!}{
\centering
\includegraphics[width=\linewidth]{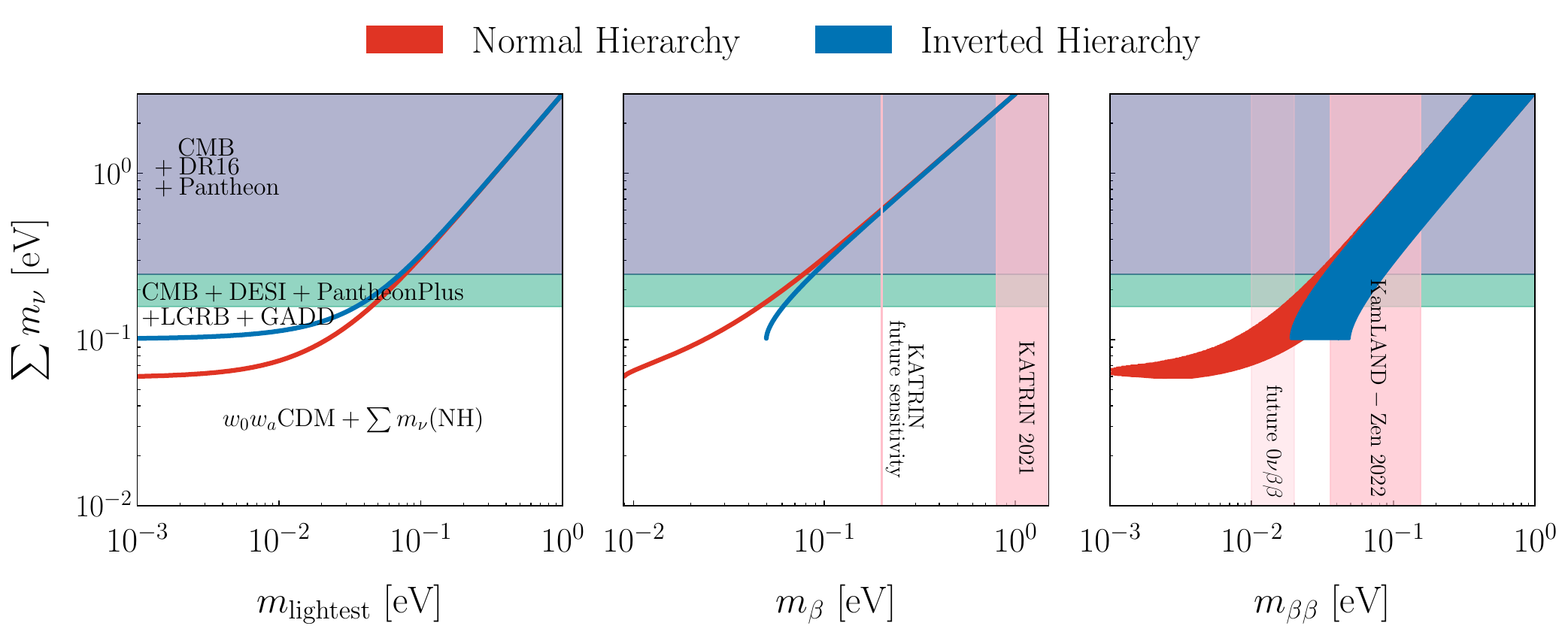}
}
\centering
\caption{Comparison of cosmological and particle physics experiment results for neutrino masses. The parameters of the neutrinos used for the mapping are taken from Refs.~\cite{dayabaycollaboration2022precision, Esteban_2020-fig4, 2020-T2K-fig4, An_2023-DYBay-fig4}. 
{\it Left panel}: $\sum m_\nu$ as a function of the mass of the lightest neutrino $m_{\rm lightest}.$ The purple and green regions indicate the $\sum m_\nu<0.248~\rm eV$ and $\sum m_\nu<0.158~\rm eV$ exclusion regions in the $w_0w_a{\rm CDM}+\sum m_\nu~({\rm NH})$ model obtained by CMB+DR16+Pantheon and CMB+DESI+PantheonPlus+LGRB+GADD, respectively.
{\it Middle panel}: $\sum m_\nu$ as a function of $\beta$-decay effective mass $m_{\beta}.$ The pink banded area indicates the latest KATRIN result $m_{\beta}<0.8~\rm eV$, while the pink line shows the KATRIN future sensitivity $0.2~\rm eV$ \cite{aker2021direct-Katrin}. 
{\it Right panel}: $\sum m_\nu$ as a function of neutrinoless double $\beta$-decay effective mass $m_{\beta\beta}.$ The pink banded area presents the latest results $m_{\beta\beta}<0.036-0.156~\rm eV$ from the KamLAND-Zen collaboration and the light pink banded area indicates future sensitivity $m_{\beta\beta}<0.01-0.02~\rm eV$ \cite{Abe_2023-KamLAND2022, Agostini_2017-0bbfutrue}.}
\label{Fig: NH-IH}
\end{figure*}

In this section, we evaluate and contrast the cosmological constraints on neutrino masses with those obtained from $\beta$-decay and neutrinoless double $\beta$-decay experiments. Utilizing the findings from the preceding section, we illustrate the outcomes of the $w_0w_a {\rm CDM}+\sum m_\nu~(\rm NH)$ model in Fig.~\ref{Fig: NH-IH} for comparative analysis.

The latest cosmological datasets have significantly tightened the upper limit on the sum of neutrino masses. It is projected that future cosmological studies will further reduce this upper limit to approximately 0.1 eV, nearing the boundary that distinguishes NH from IH. The KATRIN $\beta$-decay experiment reported in 2021 established an upper limit of $m_{\beta}<0.8$ eV, with future measurements aiming for a precision of 0.2 eV. Current cosmological assessments suggest a more stringent limit, with $m_{\beta}< 0.046$ eV for NH, which not only surpasses the precision of $\beta$-decay experiments by a factor of ten but also exceeds their sensitivity.

Moreover, the KamLAND-Zen collaboration's most recent findings on neutrinoless double $\beta$-decay, published in 2022, set an upper limit of $m_{\beta\beta}<0.036-0.156$ eV \cite{Abe_2023-KamLAND2022}. In comparison, cosmological data provide slightly tighter constraints, equivalent to $m_{\beta\beta}<0.016-0.045$ eV for NH. As a result, cosmology is emerging as a potent tool for probing neutrino mass, capable of providing independent validation of results from particle physics experiments. 

\section{Conclusion}\label{sec4}

In this paper, we employ the latest cosmological datasets, including those from DESI, DESY5, CMB observations, and additional distance measurements, to constrain the neutrino mass within the $\Lambda\rm CDM$, $w\rm CDM$, and $w_0w_a\rm CDM$ cosmological models. We explore the impacts of dark-energy EoS on measuring the neutrino mass. Utilizing Bayes factors, we also aim to identify the most likely dark energy model and the neutrino mass hierarchy.

In our analysis, we determine that the upper limit of the total neutrino mass is significantly influenced by the EoS of dark energy, which is highly consistent with the previous studies. Furthermore, the recent cosmological data have notably reduced the upper limit of the neutrino mass. Specifically, within the \( w_0w_a\rm CDM \) model, the combination of CMB, DESI, and DESY5 data yields the neutrino mass upper limits of $\sum m_{\nu} < 0.171$ eV for DH, 0.204 eV for NH, and 0.220 eV for IH. The corresponding values for the EoS parameter $w_0$ are $-0.725^{+0.066}_{-0.074}$, $-0.709^{+0.072}_{-0.072}$, and $-0.698^{+0.066}_{-0.076}$, and for $w_a$ are $-1.08^{+0.38}_{-0.30}$, $-1.19^{+0.37}_{-0.32}$, and $-1.28^{+0.38}_{-0.30}$, respectively.

The PantheonPlus dataset offers a more stringent upper limit for $\sum m_{\nu}$ than DESY5, which is closely associated with the preference of DESY5 for a more dynamically evolving dark-energy EoS. Moreover, the addition of two extra distance datasets, LGRB and GADD, results in $w_0$ values of $-0.698^{+0.069}_{-0.069}$, $ -0.680^{+0.070}_{-0.070}$, and $-0.667^{+0.068}_{-0.068}$, and $w_a$ values of $-1.28^{+0.35}_{-0.29}$, $ -1.41^{+0.34}_{-0.29}$, and $-1.52^{+0.34}_{-0.30}$, for DH, NH, and IH, respectively. This suggests that both the consideration of the neutrino hierarchy and the inclusion of further data lead to a significant deviation of dark energy from the $\Lambda$CDM model, with the latter even exceeding a $4\sigma$ deviation.

The incorporation of the LGRB and GADD datasets also effectively breaks the parameter degeneracies, enhancing the precision of the Hubble constant $H_0$ measurement by approximately $20\%$ and further reducing the upper limit of $\sum m_{\nu}.$ Furthermore, the current cosmological constraints on the neutrino mass significantly exceed those from $\beta$-decay experiments and are also more precise than those derived from current neutrinoless double $\beta$-decay experiments.

Incorporating the prior on neutrino hierarchy from particle physics experiments, the combined cosmological datasets of CMB, DESI, DESY5, LGRB, and GADD provide strong support for the $w_0w_a$CDM model with NH. The PantheonPlus dataset, while more adept at distinguishing between neutrino hierarchies than DESY5, is less effective in differentiating between dark energy models, as evidenced by the parameter constraint outcomes. Despite this, the PantheonPlus dataset still offers moderate to strong evidence in favor of the $w_0w_a$CDM model with NH.

Thus, cosmological observations and particle physics experiments mutually reinforce the study of neutrinos. The potential of future cosmological data to further refine the measurement of neutrino masses, while also distinguishing between dark energy models and neutrino hierarchies, holds great promise and is highly anticipated for the scientific community.

\begin{acknowledgments}
We are grateful to Jun-Da Pan for helpful discussions. 
This work was supported by the National SKA Program of China (Grants Nos. 2022SKA0110200 and 2022SKA0110203), the National Natural Science Foundation of China (Grants Nos. 12473001, 11975072, 11875102, and 11835009), and the National 111 Project (Grant No. B16009).
\end{acknowledgments}

\textbf{Data Availability Statement} This work has no associated data. [Author’s comment: The paper has no generated data, as in our analysis we have used CMB (Planck, ACT), DESI BAO, Pantheon (PantheonPlus, DESY5), LGRB, and GADD datasets, thereby the source of each of them has been cited in Sect.~\ref{sec2.2}.]

\textbf{Code Availability Statement} This work has no associated code/software. [Author’s comment: In this work, we have not generated any software/code. We employed the Cobaya package, interfaced with the CAMB code, to perform the MCMC method for estimating the best-fit model parameters. The relevant references are provided in Sect.~\ref{sec2.3}.]

\bibliography{nu_desi_epjc}

\begin{figure*}[htbp]
\resizebox{\textwidth}{!}{
\begin{minipage}{\textwidth}
\centering
\includegraphics[width=0.7\linewidth]{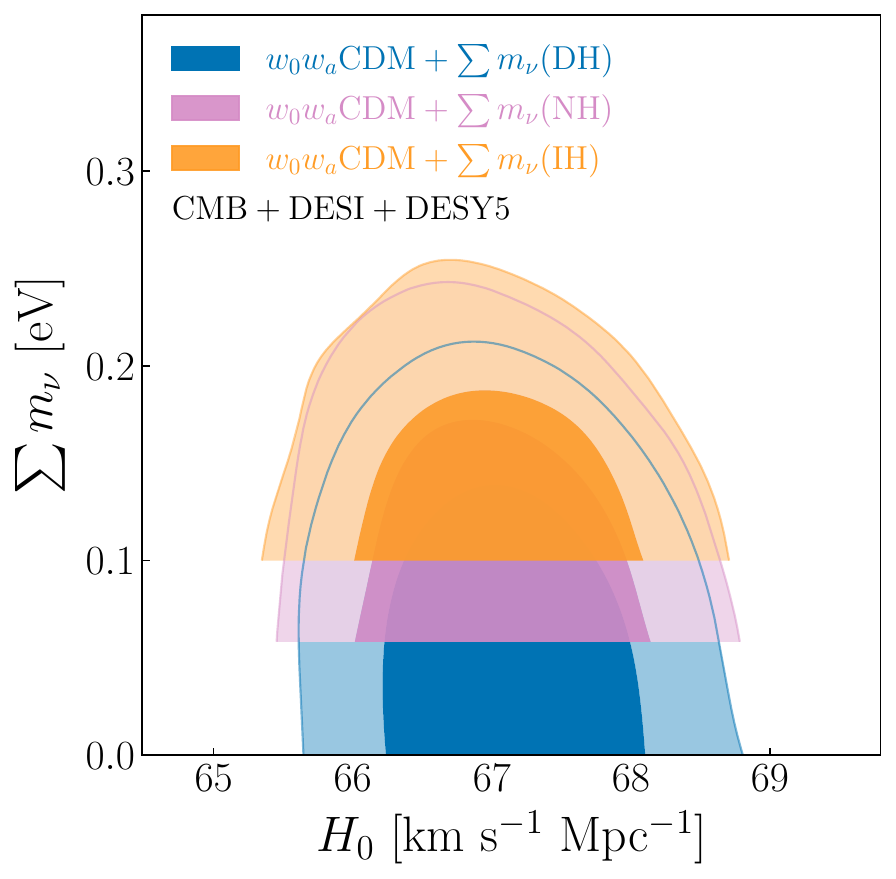}
\end{minipage}
\begin{minipage}{\textwidth}
\centering
\includegraphics[width=0.7\linewidth]{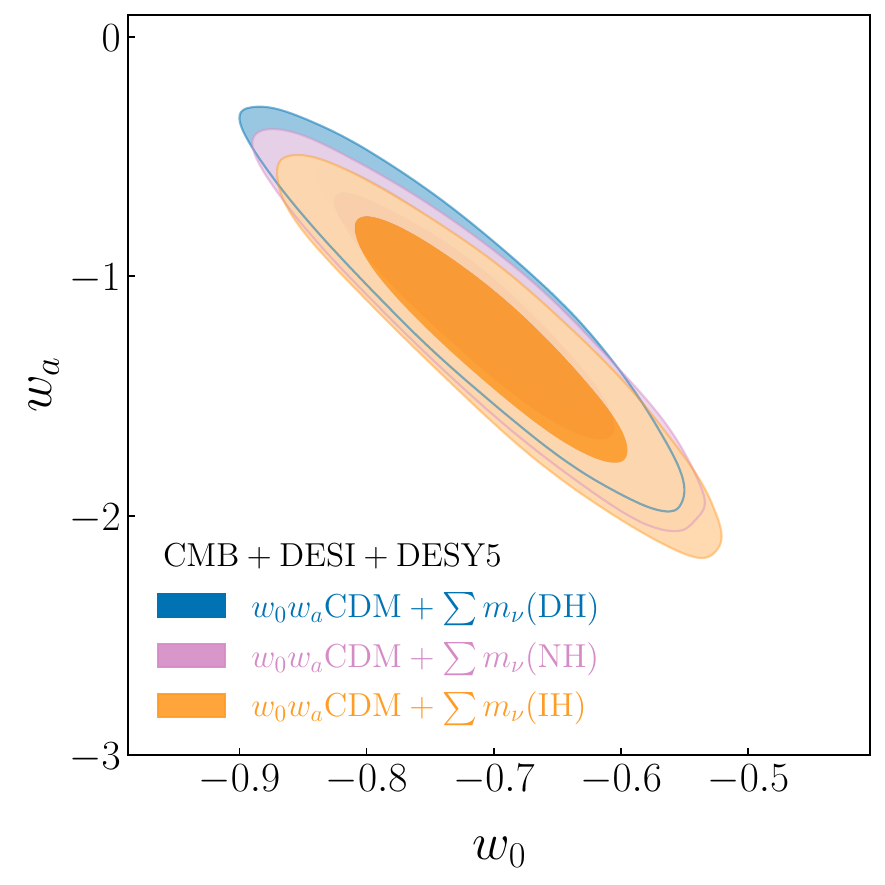}
\end{minipage}
}
\centering
\caption{The 68.3\% and 95.4\% credible-interval contours for $\sum m_\nu$ and $H_0$, as well as $w_0$ and $w_a$ in the $w_0w_a{\rm CDM}+\sum m_\nu$ models with DH, NH, and IH. The results are based on the combined CMB+DESI+DESY5 dataset.}
\label{fig5}
\end{figure*}
\appendix
\section{} \label{sec:appendix-hierarchy}

To provide a direct visual comparison of the effects of different neutrino mass hierarchies, we present Fig.~\ref{fig5}, which shows the credible-interval contours of $\sum m_\nu$ and $H_0$, as well as $w_0$ and $w_a$ in the $w_0w_a{\rm CDM}+\sum m_\nu$ models with DH, NH, and IH. This analysis is performed using the combined CMB+DESI+DESY5 dataset as a representative case. As illustrated in Fig.~\ref{fig5}, the variations in $H_0$, $w_0$, and $w_a$ across the three mass hierarchy scenarios are statistically nonsignificant. The results show that the choice of neutrino mass hierarchies does not significantly influence the inferred cosmological parameters.

\end{document}